\documentclass[sigconf, 10pt]{acmart}\usepackage[]{graphicx}\usepackage[]{color}
\makeatletter
\def\maxwidth{ %
  \ifdim\Gin@nat@width>\linewidth
    \linewidth
  \else
    \Gin@nat@width
  \fi
}
\makeatother

\definecolor{fgcolor}{rgb}{0.345, 0.345, 0.345}
\newcommand{\hlkwb}[1]{\textcolor[rgb]{0.69,0.353,0.396}{#1}}%

\usepackage{framed}
\makeatletter
 {\par\unskip\endMakeFramed%
 \at@end@of@kframe}
\makeatother

\definecolor{shadecolor}{rgb}{.97, .97, .97}
\definecolor{messagecolor}{rgb}{0, 0, 0}
\definecolor{warningcolor}{rgb}{1, 0, 1}
\definecolor{errorcolor}{rgb}{1, 0, 0}
\usepackage{alltt}
\AtBeginDocument{%
  \providecommand\BibTeX{{%
    \normalfont B\kern-0.5em{\scshape i\kern-0.25em b}\kern-0.8em\TeX}}}

\usepackage{array}
\newcolumntype{L}[1]{>{\raggedright\let\newline\\\arraybackslash\hspace{0pt}}m{#1}}

\IfFileExists{upquote.sty}{\usepackage{upquote}}{}
\begin{document}

\title{The impact of overbooking on a pre-trial risk assessment tool}

\author{Kristian Lum}
\email{kl@hrdag.org}
\affiliation{
\institution{Human Rights Data Analysis Group}}

\author{Chesa Boudin}
\email{chesa.boudin@sfpd.gov}
\affiliation{%
  \institution{San Francisco Public Defender's Office}
 % \streetaddress{P.O. Box 1212}
}

\author{Megan Price}
\email{meganp@hrdag.org}
\affiliation{
\institution{Human Rights Data Analysis Group}}

\renewcommand{\shortauthors}{Lum, Boudin, and Price}

\begin{abstract}
Pre-trial risk assessment tools are used to make recommendations to judges about appropriate conditions of pre-trial supervision for people who have been arrested. Increasingly, there is concern about whether these models are operating fairly, including concerns about whether the models' input factors are fair measures of one's criminal activity. In this paper, we assess the impact of booking charges that do not result in a conviction on a popular risk assessment tool, the Arnold Public Safety Assessment. Using data from a pilot run of the tool in San Francisco, CA, we find that booking charges that do not result in a conviction (i.e. charges that are dropped or end in an acquittal) increased the recommended level of pre-trial supervision in around 27\% of cases evaluated by the tool. %We also find that, while unsubstantiated booking charges lead to charge-based exclusions more often for Black arrestees, this does not lead to a large racial disparity in the impact of overbooking on the ultimate recommendations-- as a larger proportion of Black arrestees are already in the highest risk group irrespective of charge-based overrides. 

\end{abstract}

%\begin{CCSXML}
%<ccs2012>
%  <concept_id>10010520.10010553.10010562</concept_id>
%  <concept_desc>Computer systems organization~Embedded systems</concept_desc>
%  <concept_significance>500</concept_significance>
%  <concept_id>10010520.10010575.10010755</concept_id>
%  <concept_desc>Computer systems organization~Redundancy</concept_desc>
%  <concept_significance>300</concept_significance>
%  <concept_id>10010520.10010553.10010554</concept_id>
%  <concept_desc>Computer systems organization~Robotics</concept_desc>
%  <concept_significance>100</concept_significance>
%  <concept_id>10003033.10003083.10003095</concept_id>
%  <concept_desc>Networks~Network reliability</concept_desc>
%  <concept_significance>100</concept_significance>
%</ccs2012>
%\end{CCSXML}
%
%\ccsdesc[500]{Computer systems organization~Embedded systems}

\setcopyright{none}
\settopmatter{printacmref=false}
\renewcommand\footnotetextcopyrightpermission[1]{} 
\pagestyle{plain}

\keywords{risk assessment, police accountability, overbooking, fairness}

\maketitle

\section{Introduction}
When a person is arrested a decision is made as to whether that person should be released before their case has concluded and, if so, under what conditions. Such decisions take many factors into account, including the risks of releasing that person back into the community. Actuarial risk assessment-- statistical models that output a defendant's estimated risk (often risk of re-arrest or risk of failure to appear for court)-- are increasingly used to aid decision-makers in this highly consequential decision. 

Recently, concerns have been raised about the fairness of risk assessment models in the criminal justice context, particularly around race-based discrimination \citep{angwin2016machine}. Much of the technical discussion in this area has centered around defining the fairness of a model in terms of racial parity along several, potentially conflicting, measures of predictive performance \citep{skeem2016risk, dieterich2016compas, berk2018fairness, chouldechova2017fair, Kleinberg:2018:ITA:3219617.3219634, Solow-Niederman:2019aa}. Other concerns about risk assessment focus on the (un)fairness of the inputs to the models \citep{johndrow2019algorithm, eckhouse2019layers}. For example, \cite{harcourt2015risk} has argued that the use of criminal history in risk assessment serves as a proxy for race and predictive models that rely upon this factor serve to justify the continued over-incarceration of minorities. \cite{Mayson:2019aa} offers a discussion of many of the arguments supporting and refuting the fairness of risk assessment in criminal justice.

In this paper, we explore the ``fairness" of one particular input of a popular risk assessment tool. Using data provided by the San Francisco Public Defender's Office (SFPD), we evaluate how often ``overbooking"--- booking an arrested person under charges that are higher than may be warranted by the facts of the case--- influences  Arnold Ventures'\footnote{Formerly the Laura and John Arnold Foundation} Public Safety Assessment (PSA). Specifically, using data from a pilot run of the model in mid 2016 to mid 2017 in San Francisco, CA, we investigate how often charges that are ultimately unsubstantiated by the courts cause the PSA to make more restrictive recommendations than it would have if it had only used charges for which the person was ultimately found guilty. Allowing unsubstantiated charges to increase one's pre-trial supervision recommendations %, thereby reducing their chances of pre-trial release if judges heed the model's recommendations,
 runs counter to an intuitive notion of fairness in the sense that charges that are ultimately dropped or not sufficiently supported by the facts of the case to result in a conviction ought not to be used to justify more restrictive release conditions. Thus, this work assesses the fairness of the model both from the point of view of the inputs as well as their effects on the predictions and recommendations of the tool. To our knowledge, this is the first study of its kind to analyze the effect of charge-based inputs to risk assessment in this way. 

In Section \ref{sec:overbooking} we give context for and define overbooking for the purposes of our analysis. Section \ref{sec:PSA} gives a description of the risk assessment model we study here, the PSA,  including a discussion of how booking charges are factored into the ultimate recommendations of the model. In Section \ref{sec:Data}, we describe the data we used in the analysis, and in Section \ref{sec:Analysis}, we describe how we analyzed the data. Section \ref{sec:Results} presents our findings, Section \ref{sec:Limitations} discusses the limitations of the the work, and Section \ref{sec:conclusion} concludes.

\section{Overbooking} \label{sec:overbooking}
We assess the impact of overbooking on the PSA in San Francisco, CA over several months in 2016 and 2017. The local context here is important. According to a recent report based on an examination of cases taken on by the SFPD \cite{owens2017examining}, ``[p]eople of color receive more serious charges at the initial booking stage, reflecting decisions made by officers of the San Francisco Police Department.'' They concluded that the disparities in booking charges were one of the largest factors contributing to racial disparities in San Francisco's criminal justice system as a whole. 

In San Francisco and elsewhere, booking officers (the police officers who formally file the initial charge) have a high degree of discretion when determining the booking charge. Specifically, when a law enforcement officer in the field makes an arrest they can either cite and release the arrestee (for certain low level-misdemeanors), or book the arrestee into county jail. When the arresting law enforcement agency transfers custody of the arrestee to the Sheriff (who runs the jail), it must provide the legal basis for the person’s incarceration: the booking charge(s). Thus, typically, the officer who delivers the arrestee to the jail tells the booking deputy at the jail on what charges they are booked. There is virtually no oversight or feedback mechanism on the discretion to choose booking charges. So, for example, in a case where a person is arrested with a single illegal firearm, the arresting officer could choose to book the person on charges for three separate guns (or, in an extreme case, even murder).\footnote{While this sounds extreme, this example is based on one of the authors' experience as a public defender.} In reviewing the case, the District Attorney’s office should, in theory, discharge any inflated charges and rebook the arrestee on charges which are actually based in the evidence, but by that stage the risk assessment algorithm has already generated its recommendation using the booking charges as an input. 

Many tools, like the PSA, nevertheless rely on the pre-conviction charges as an input because of the need to generate risk reports quickly. There is typically a delay of approximately 48 hours\footnote{Penal Code section 825} between the time of booking into jail and the District Attorney’s decision about whether and what files to charge (rebooking). One of the goals of incorporating risk assessment into the pre-trial process is to expedite the release of low-risk arrestees. If the algorithms could not be run until the district attorney had made a rebooking decision, everyone would have to wait in jail for days. Thus, to effectuate speedy release for some, the PSA relies on the police booking charge to generate reports within 24 hours, while the district attorney is still considering rebooking. 

The ultimate objective of this analysis is to assess how often ``unfair" booking charges caused the PSA to recommend excessively restrictive conditions of pre-trial supervision. To do this, we define unfair booking charges to be those charges associated with the PSA that do not go on to result in a conviction.  Implicitly, this  assumes that the conviction charges (the charges to which the person pleads or is found guilty) are a fair and accurate representation of the person's criminal activity. We acknowledge that this measure of the true severity of the crimes is imperfect.

On the one hand, there may exist cases where a defendant did commit the crime on which they are booked, and the defense just barely establishes reasonable doubt leading to an acquittal for those charges. It may seem disingenuous to label such charges unfair or unsubstantiated. On the other hand, as discussed in \cite{kurlychek2019cumulative} and the many citations therein, ``initial appraisals of dangerousness and culpability may  send signals to later system actors, setting into motion a dynamic pattern of cascading disadvantage." For example, the rebooking charges are often identical to the booking charges. The rebooking charges then define the starting place from which plea deals are negotiated. Higher booking charges may follow the defendant through the whole plea process, with the plea deals they are offered being anchored to the initial charges, reducing the chances that they are offered deals with less severe charges. 

Relatedly, many recent studies have found that pre-trial detention {\it causes} defendants to accept guilty pleas to charges that they otherwise would not have been convicted of had they been free pre-trial \cite{gupta2016heavy, leslie2017unintended, lum2017causal, stevenson2018distortion, dobbie2018effects}.  To the extent that the risk assessment is heeded by judges, booking charges that cause the tool to make more restrictive supervision recommendations might then indirectly be causing those charges to be ``substantiated" through a guilty plea. Although conviction charges are an imperfect ground truth in that they likely exhibit similar (though hopefully less severe) biases to the booking charges, we believe this is the best measure of appropriate charges possible given the data we have available. 

%If anything, we believe this comparison offers a lower bound on the effect of overbooking on the risk assessment model, as 

%unfounded booking charges that are passed through a risk assessment to influence a judge to order more restrictive conditions of release or deny release altogether may {\it cause} the arrestee to ultimately be convicted of charges that would otherwise be dismissed \citep{leslie2017unintended, gupta2016heavy, stevenson2018distortion, lum2017causal, dobbie2018effects}.%With regard to the connection between overbooking and racial disparities, as summarized by \cite{owens2017examining},``booking decisions influence downstream decisions made by district attorneys, public defenders, and judges. District attorneys and public defenders are making what appear to be race-neutral decisions in response to the charges brought to them by the police – but police bring more severe charges against Blacks and Latinx relative to Whites, and that then persists throughout the case adjudication process."  %Likely a lower bound on teh effect

\section{The PSA}\label{sec:PSA}
The PSA developed by Arnold Ventures is a popular pre-trial risk assessment tool that is intended to ``reduce the burden placed on vulnerable populations at the forefront of the criminal justice process."  \citep{demichele2018public} It has been adopted by more than 40 jurisdictions around the country \citep{Redcross:2019aa}. The purpose of the tool is to make recommendations about the appropriate level of pre-trial supervision for people who have been arrested. 

%add something about the importance of understanding this
In this section, we describe the PSA as it was administered during the period of our study.  Since then, Arnold Ventures has modified its terminology for some of the components of the PSA as well as some of the procedures for translating the risk scores to recommendations.\footnote{As we describe each component of the version of the PSA as it existed during the period under study, we will highlight analogues to each component under the new version of the PSA.}
%\footnote{ In the version of the PSA we report below, the process by the booking charges enter the calculation is delineated in steps, with steps two and four, in particular, presenting the opportunity for increases in levels of pre-trial supervision as a result of the bookings charges. In the new version of the PSA, these charge-based increases are not framed as steps in the process but rather as points for ``additional guidance" with an example of additional guidance being ``If the current charge is a first- or second-degree violent felony, the person may be placed on Release Level 3 (the highest release level), regardless of the PSA scores" (an analogue of a step 2 exclusion, as described below) }  
However, it is not mandatory that jurisdictions adopt the revised version, and many jurisdictions remain using a version similar to that described below.  In fact, the version of the PSA described here is still in use in San Francisco.

In order to create a risk profile for a newly arrested person and pre-trial supervision recommendation, the tool combines several separate predictions: risk of failure to appear at a future court date (FTA), risk of re-arrest for new criminal activity (NCA), and  risk of re-arrest for new {\it violent} criminal activity (NVCA). 

Each of the predictions are calculated as a function of some subset of the following:
\begin{itemize}
\item age at current arrest, 
\item whether there are pending charges at the time of the current offense, 
\item whether the arrested person has any prior misdemeanor convictions, 
\item whether the arrested person has any prior felony convictions, 
\item whether the arrested person has any prior convictions (misdemeanor or felony), 
\item the number of prior violent convictions, 
\item the number of prior failures to appear for court dates in the past two years, 
\item whether the person failed to appear prior to two years before the offense,  
\item whether the individual has been incarcerated as the result of a conviction in the past, 
\item whether the current booked offense is considered violent. 
\end{itemize}

Violent charges are determined by inclusion on the extensive PSA Violent Offense list for California.  We include a representative list of charges that appear on the violent offense list in Table \ref{tab:violent-offenses} in the appendix. Both FTA and NCA predictions are conveyed on a six point scale, with higher values corresponding to a higher predicted likelihood of the corresponding undesirable outcome. The NVCA prediction is a binary prediction, with a value of one (sometimes called a NVCA flag or violence flag) corresponding to a higher predicted likelihood of future arrest for a violent crime. Each of these predictions are calculated as a linear combination of integer valued weights. This information and information on the weights associated with each factor for each type of prediction as well as extensive documentation around the implementation of the tool is available on Arnold Ventures portal\footnote{https://www.psapretrial.org/about/factors.} in the Guide to the Release Conditions Matrix \citep{Ventures:aa}.

Calculating the FTA, NCA, and NVCA sub-scores constitutes step (1) of a four-step process. In the subsequent three steps, these risk scores are combined with the booking charges using pre-defined sets of rules to transform the risk scores into recommendations.  

 In step (2) of the PSA, three determinations are made:

\begin{itemize}
\item whether the person was extradited for the current booked offense;
\item whether current booked offense is among the following offenses or is a conspiracy, attempt, solicitation, or FTA of any for those offenses;

\begin{itemize}
\item Murder
\item Voluntary Manslaughter
\item Aggravated Mayhem
\item Torture
\item Felony Sexual Assault
\item Robbery
\item Carjacking
\item Felony Domestic Violence
\item Felony Stalking
\item Violation of a Domestic Violence Protective Order 
\item Escape
\end{itemize} 

\item whether the current booked offense is deemed violent according to the California PSA List of Violent Offenses and the NVCA flag calculated in step (1) is indicated.  
\end{itemize}

 If any of these conditions are true, the individual is automatically given a ``Release Not Recommended" (the most restrictive possible recommendation) and the assessment need not continue. This is called a ``charge-based exclusion," and we refer to charges that trigger a charge-based exclusion as ``exclusion charges." \footnote{Under the new version of the PSA, there is no longer a charge-based exclusion as a formal step in the process. The analogue to charge-based exclusions now fall under ``additional guidance." For example, in a sample PSA given in \cite{Ventures:aa}, the guidebook gives an example of additional guidance to augment the PSA as ``If the current charge is a first- or second-degree violent felony, the person
may be placed on Release Level 3 (the highest release level), regardless of
the PSA scores." This is analogous to a step (2) exclusion under the version of the PSA described here.}
 If no charge-based exclusion is made, the assessment continues to step (3).

\begin{figure}[h]
\includegraphics[width=3.65in]{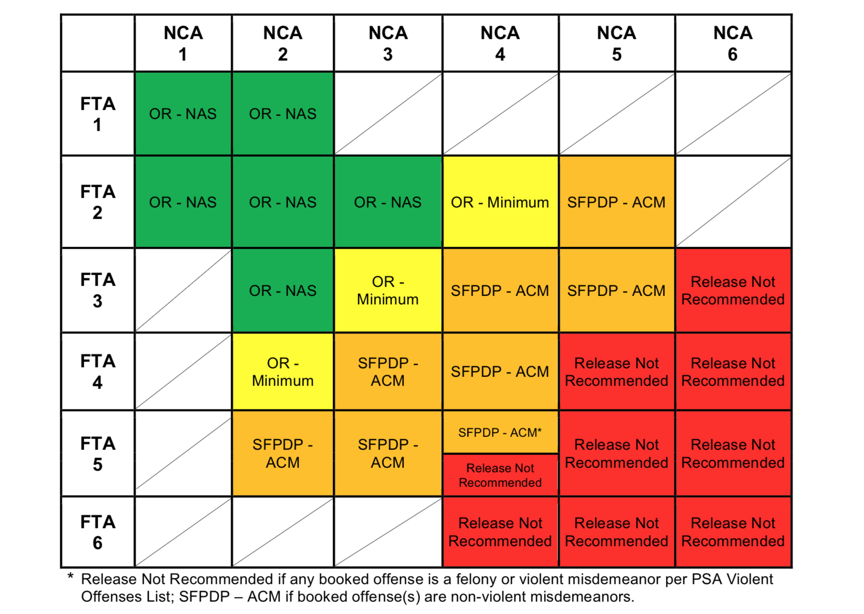}
\caption{\label{fig:dmf} Decision Making Framework used in San Francisco pilot study. This matrix translates the empirical risk estimates of NCA and FTA into policy recommendations about the appropriate level of pre-trial supervision for a person with that estimated risk profile.}
\end{figure}

 In step (3), the six point NCA and FTA predictions are combined using the Decision Making Framework (DMF) shown in Figure \ref{fig:dmf} to arrive at what we refer to as an ``initial recommendation."\footnote{This terminology may be confusing, as this comes after step (2) so chronologically in some sense is not initial. By this we mean to say that this is the recommendation that would be given without considering any charge-based amendments to the recommendation. This can be calculated even if an individual has a charge-based exclusion. } Each cell in the matrix corresponds to a recommended level of supervision. For example, if an individual has an FTA prediction of 2 and a NCA prediction of 3, one would find the entry in the DMF in the second row and third column to arrive at an initial recommendation of OR-NAS, the lowest recommended level of supervision possible.  
  
In this implementation of the tool, the possible levels of supervision in the DMF in order of increasing level of supervision were: (1) No Active Supervision-- release on own recognizance and receive court reminders (OR-NAS); (2) Minimum Supervision-- release on own recognizance with court reminders and twice weekly phone reporting (OR-Minimum); (3) Assertive Case Management-- release on own recognizance or under supervision, including court date reminders, four times weekly reporting with two to four of those times reporting in person, and an out of custody needs assessment (SFPDP-ACM); and, (4) Release not Recommended.\footnote{What is referred to as the decision-making framework (DMF) in this section has a direct analogue in the new version of the PSA called the Release Conditions Matrix (RCM). One important differentiator between the versions is that in the version described here, the DMF may recommend pre-trial detention. Under the new version with the RCM, according to \cite{Ventures:aa} ``Detention is not included in the matrix because eligibility for detention is based on state law, and the matrix becomes relevant only after a judicial officer decides a person will be released."}

%This mapping from risk scores to recommendations is jurisdiction-specific and is determined at the local level. This local tailoring of which particular booking charges trigger increases to the level of supervision arises partially out of necessity (charge codes, for example, vary across jurisdictions) and partially so that the local considerations and norms relevant to those implementing the PSA can be incorporated into the recommendation process. 
% 
Finally, in step (4), two more determinations are made:

\begin{itemize}
\item whether the current booked offense is among the following charges or the current booked offense is a solicitation, conspiracy, attempt or FTA for any of the charges on that list; 
\begin{itemize}
\item Violation of other Protective Orders
\item Person to Person Sex Crime
\item Arson
\item Involved the \underline{Use} of a Weapon, Caustic Chemical, Flammable Substance, or Explosive,
\item Felony inflicting Great Bodily Injury
\item Misdemeanor Domestic Violence
\item Misdemeanor Stalking
\end{itemize}

\item  whether the current booked charge is not on the list of violent offenses but the NVCA flag was triggered.
\end{itemize}

\noindent If either of these conditions are true, the initial recommendation is increased one level. For example, if the initial recommendation was OR-NAS, it is increased to OR-Minimum. This recommendation is then the final pre-trial supervision recommendation. \footnote{Similar to step (2), step (4) increases are now included under ``additional guidance" rather than as a formalized step. For example, in the sample PSA given in \cite{Ventures:aa}, ``If there is an NVCA flag, consider increasing the person’s release level by one level" is given as an example of additional guidance that could be included. This is analogous to the step (4) increase given here, though less formalized.} We refer to this as a ``charge-based bump-up" and charges that trigger a charge-based bump-up as ``bump-up charges."

In summary, abandoning jargon and simplifying to the extent possible, we conceptualize the process as follows: an individual's initial recommended level of supervision is determined by combining predictions about their likelihood of FTA and NCA (predictions that do not rely on charge-based information). If there are very serious booking charges (i.e. exclusion charges) or the booking charges are violent and the person is predicted to have a higher likelihood of re-arrest for a violent crime, then the individual is automatically recommended for the highest level of supervision. If there are moderately serious booking charges (i.e. bump-up charges) or the individual is predicted to have a higher likelihood of re-arrest for a violent crime despite the current charges not being violent, then the initial recommendation is increased to the next highest level. 

\section{Data} \label{sec:Data}
In collaboration with the SFPD, we obtained data collected during San Francisco's pilot study of the PSA. This study was conducted over the period from mid-2016 to mid-2017.  During this time period, the SFPD saved PSAs for their clients as scanned image files. From the SFPD, we obtained 2450 of these files, the information from which was manually entered into a spreadsheet. The following fields were collected from each PSA: the defendant's unique identification number, name, date of birth, arrest date, date on which the PSA was conducted, NVCA prediction, NCA prediction, FTA prediction, a list of booked charges and corresponding charge codes, the recommendation of the PSA, an indicator of whether the recommendation was the result of charge-based exclusion, and an indicator of whether a charge-based bump-up was applied to the recommendation.  We also recorded several of the individual risk factors listed on the PSA for each defendant: age at current offense, prior conviction (misdemeanor or felony), and prior violent conviction. %Combined with the current charge information, these are the risk factors necessary to calculate the NVCA flag.

We separately obtained data from San Francisco's court databases on all defendants who interacted with the court system during the time period of the pilot program. %This dataset consists of 13182 defendants with a total of 62364 charges. 
For each defendant, we obtained a unique identification number, name, date of birth, list of booking charges, list of charges filed by the district attorney, and the disposition code for each individual charge. The disposition codes define whether each charge resulted in a conviction and are the basis on which we retrospectively calculate which charges were substantiated and which were not. 

\section{Analysis} \label{sec:Analysis}

Our analysis consists of three parts: de-duplication and record-linkage; validation; and counterfactual analysis. The goal of the de-duplication and record linkage is to ensure that each arrest only appears once in our dataset and that each arrest record in the PSA data is linked to the correct court case in the court data. This linkage allows us to see whether the booking charges included in each PSA ultimately resulted in a conviction, information that is only contained in the court records. It also provides us additional demographic information about the people who were assessed by the PSA. The result of this step is one unified dataset that contains one copy of each PSA that was administered during the pilot program, the outcome of all charges associated with each PSA, and additional demographic information about the individual to whom each PSA pertains. In  completing this process, we drop 64 records due to incompleteness, 419 records as duplicates\footnote{Many of the dropped records pertain to the same few arrests, which it seems were each saved multiple times.}, and 31 records due to an inability to establish a definitive match in the court data. In the end, this leaves us with 1916 records with which to do the analysis. Details of the decision rules used to de-duplicate and match are found in Section \ref{sec:matching} of the appendix. 

In the validation phase, we create PSA-reproduction code and verify that our code accurately reproduces the outputs of the human-administered PSA when given the same inputs. To do this, we apply our PSA-reproduction code to the linked dataset described above. This code takes the FTA and NCA predictions\footnote{Because the FTA and NCA predictions do not depend on booking charge, these are held constant and we need not re-calculate them. That is, we do not need to verify that we can reproduce them, as this analysis pertains only to effects driven by perturbations of the booking charges, and these predictions do not change as a function of the booking charges.}, several variables measuring criminal history, and the booking charges as listed in the court records. It then outputs each of the components of the PSA: a charge-based exclusion indicator, a charge-based bump-up indicator, a violence flag, and a final recommendation. We then compare each of these outputs to those same components as recorded on the original PSA forms. A high rate of agreement assures that our code is accurately representing the PSA.

We find that our calculation of the NVCA flag agrees with that listed on the PSA form for 99.3\% of the cases in our dataset. With regard to charge-based exclusions, we find that our reproduction is in agreement with the original PSA data for 99.4\% of the records. In the PSA, calculating the charge-based bump-up is not necessary if an exclusion is determined because the recommended level of supervision cannot further be increased. For this reason, we compare our reproduction of charge-based bump-ups to the PSA data only for those records for which a charge-based exclusion was not indicated in the original PSA data. We find that our reproduction is in agreement with the original administration of the PSA for 99.7\% of these records. Finally, we compare our calculation of the final recommendation to the recommendation given in the PSA data. We find an agreement of 97.4\%, which is slightly lower than for the other components. Nearly all of the disagreements occurred for individuals who fell within the one specific cell of the DMF which required additional determinations to be made (an FTA of 5 and a NCA of 4, shown as the split cell in Figure \ref{fig:dmf}). Based on a manual review, we believe there may have been differing interpretations by the staff administering the PSA as to how these determinations should be applied.  In any case, for those cases that do not fall into this cell of the DMF, our reproduction of the final recommendation is in agreement with that listed on the PSA form 99.5\% of the time. A full discussion of the validation process is given in Section \ref{sec:validation} of the appendix. 

Finally, having verified that our code accurately reproduces the PSA, we apply our PSA-reproduction code to a counterfactual scenario in which only the charges that resulted in a conviction are used in the calculation of the PSA. This results in two sets of calculations to compare: (a) the PSA's recommendation (and each of its components) based on the booking charges, and (b) the PSA's recommendation (and each of its components) based only on conviction charges. %In some cases, additional charges were later added after booking, so it is not always the case that the conviction charges are a subset of the booking charges. 
To evaluate the impact that unsubstantiated charges had on the PSA, we compare calculations (a) and (b).

\section{Results}\label{sec:Results}
%Occasionally, court cases are slow to be resolved-- sometimes because the defendant does not re-appear for subsequent court dates, sometimes for other reasons--, and our data reflects this. In our data round(100-mnum$percent_with_some_charge_disposed*100,1)\% of the cases in the original PSA data have not  yet been disposed.  Although we would ideally have all of the dispositions, we consider this to be a suitably small percent of the data, and the remainder of the analysis pertains only to cases that have been disposed. 

%We now turn to assessing the impact of overbooking, or the effect of unsubstantiated booking charges, on the PSA. To do this, we calculate the NVCA flag, charge-based exclusions,  charge-based bump-ups, and the resulting recommendation using only the charges the defendant was convicted of rather than using all booking charges as in the original administration of the tool. 
In this section, we compare the results of the PSA calculated using the conviction charges to the results of the PSA calculated using the booking charges.\footnote{We do not compare to the original PSA results directly to isolate the effect of altering the input charges. If we compared to the original PSA components, some of the differences we identify may, in fact, be due to some of the differences in interpretation we highlighted in the validation section above.} Throughout this section, the results presented pertain only to the cases for which all charges had been disposed (or settled) at the time of the analysis, which includes 88.3\% of the records.

There is some nuance around which charges should count as convictions. For example, sometimes multiple cases are bundled into a single plea agreement. Should all charges associated with that bundle be counted as conviction charges or only those conviction charges that are part of the case originally associated with the administration of the PSA? Ultimately, we decided that only charges that pertain directly to the arrest that triggered the administration of the PSA ought to be eligible, though we acknowledge that others might disagree with this definition. Thus, for the purposes of this analysis, we define ``conviction charges" to be those charges associated with the arrest that triggered the administration of the PSA for which the arrestee was found or plead guilty.\footnote{According to the codebook we received, this is all disposition codes greater than 159.  Additionally, by manual review, we have found that cases in which the case is listed as resolved, if some charges  associated with a case number have disposition code 72 (plead guilty to other charges) and others charge codes associated with that same case number have disposition code 0, those with disposition code 0 are the ones the individual was convicted of. This was confirmed on several cases by looking at alternative sources of information available in other systems that are not in a database form amenable to statistical analysis.} 

This definition creates some situations where a case outcome indicates that the individual pleaded guilty to other charges, but none of the charges to which they pleaded guilty are associated with the original case. In this scenario, the conviction-charge-PSA is calculated as though there were no charges eligible to trigger charge-based exclusions, bump-ups, or to be considered violent, despite the fact the individual was convicted on some charges (just none that were filed as part of the case associated with their PSA form). We performed additional analysis removing all cases for which a guilty plea was indicated but the individual did not plead guilty to any of the charges associated with the original case. While the exact numbers were lower than those presented in the remainder of this section, qualitatively the results were the same.

Table \ref{tab:comparison-all-nogroup} shows the rate of charge-based exclusions, bump-ups, NVCA flags,  and the average recommendation level when calculating each of the components of the PSA. The average recommendation level is based on mapping each level of supervision to its numeric rank: the lowest level of supervision is mapped to one, the second to two, etc. The Charges column gives the charges used to calculate the PSA-- either booking charges or conviction charges. The difference between the rate calculated under the booking charges and the rate calculated under the conviction charges is also shown. To test for statistical significance between the components of the PSA under the two input charge conditions, we performed standard statistical hypothesis tests. When comparing the rate of exclusions under booking charges to the rate of exclusions under conviction charges we perform a difference of proportion test. To compare the recommendations, we performed a Wilcoxon rank-sum test, as the recommendations are ordered categorical. Statistical significance at the $\alpha < 0.001$ level is indicated by `$*$' for the differences shown in the results tables.\footnote{All p-values are significant at at least the 0.001 level, even after a Bonferroni correction for multiplicity.} We see that the rate at which charge-based exclusions,  bump-ups, and NVCA flags occur is much higher when we consider booking charges relative to when we consider conviction charges as inputs. The average level of recommended pre-trial supervision is also elevated under the calculation using the booking charges relative to that using the conviction charges.

% latex table generated in R 3.5.1 by xtable 1.8-4 package
% Wed Dec  4 11:33:45 2019
\begin{table}[ht]
\centering
\begin{tabular}{lllll}
  \hline
Charges & exclusions & bump-ups & nvca & rec \\ 
  \hline
Conviction & 8.7 & 9.1 & 9.3 & 2.5 \\ 
  Booking & 29.4 & 23.7 & 20.2 & 3 \\ 
  Difference & 20.7 * & 14.6 * & 10.8 * & 0.5 * \\ 
   \hline
\end{tabular}
\caption{Percent of cases with exclusions, bump-ups, nvca flags, and the average recommendation by input charges.} 
\label{tab:comparison-all-nogroup}
\end{table}

Table \ref{tab:comparison-all-proportion-affected} shows the proportion of people who received a charge-based exclusion, a charge-based bump-up, or an NVCA flag when the PSA was calculated using the booking charges but not when it was calculated using the conviction charges. It also shows the percent of people who received a recommendation for more restrictive conditions under the booking charges than under the conviction charges.\footnote{This is different than what is shown in Table \ref{tab:comparison-all-nogroup}, as under the former calculation, cases in which the individual was convicted of more severe charges than those under which they were booked offset cases where the reverse occurs.} 
We find that a substantial portion of the cases (nearly 30\%) would have had a lower recommended level of supervision if their PSA had been based only on the charges they were ultimately convicted of.

%remove NVCA to be consistent with previous table! For race comparison, statistical significance should be determined by difference between groups? 
% latex table generated in R 3.5.1 by xtable 1.8-4 package
% Wed Dec  4 11:33:45 2019
\begin{table}[ht]
\centering
\begin{tabular}{rrrr}
  \hline
exclusions & bump-ups & nvca & rec \\ 
  \hline
20.9 & 17.0 & 10.9 & 27.4 \\ 
   \hline
\end{tabular}
\caption{Percent of cases for which each PSA component was higher under the booking charges than under the conviction charges.} 
\label{tab:comparison-all-proportion-affected}
\end{table}

%\subsection{Assessing the impact of overbooking by race}
Next we turn to understanding whether overbooking's effect on the PSA differs by race group. For this analysis, we disaggregate the data into two race categories: Black and non-Black. This is an obvious over-simplification, as is any racial categorization. However, based on our analysis of the consistency of racial classification within the court data, we have determined this categorization scheme introduces the fewest problems with inconsistent classification. A full discussion of how we arrived at this decision is available in the appendix. 

Table \ref{tab:comparison-all-by-race} shows equivalent quantities to those shown in Table \ref{tab:comparison-all-nogroup}, now disaggregated into the two race groups. We find that overbooking had a larger impact on the rate of charge-based exclusions and the assignment of the NVCA flag for  Black people than non-Black people in this data. It had a larger impact on charge-based bump-ups on the non-Black population. However, the impact of overbooking on the ultimate recommendation is, roughly speaking, similar between the two groups. The proportion of cases for which charge-based overrides, bump-ups, NVCA flags were triggered or the recommendation was higher under the conviction charges than under the booking charges is given in Table \ref{tab:all-comparison-prop-race} disaggregated by race. Under this summary of the data, we again see that unsubstantiated charges led to charge-based exclusions at a higher rate for Black defendants than non-Black defendants. However,  there is little difference between the groups in terms of the impact of unsubstantiated charges on charge-based bump-ups or on the final recommendation.

\begin{small}% latex table generated in R 3.5.1 by xtable 1.8-4 package
% Wed Dec  4 11:33:45 2019
\begin{table}[ht]
\centering
\begin{tabular}{llllll}
  \hline
Charges & group & exclusions & bump-ups & nvca & rec \\ 
  \hline
Conviction & non-black & 7.9 & 8.2 & 7.7 & 2.4 \\ 
  Conviction & black & 9.7 & 10.3 & 11.5 & 2.6 \\ 
  Booking & non-black & 25.9 & 23.2 & 15.9 & 2.9 \\ 
  Booking & black & 33.9 & 24.3 & 25.7 & 3.1 \\ 
  Difference & non-black & 18 * & 15 * & 8.2 * & 0.5 * \\ 
  Difference & black & 24.2 * & 14.1 * & 14.2 * & 0.5 * \\ 
   \hline
\end{tabular}
\caption{Comparison of effects of charges on components of PSA by race for all disposed cases in data} 
\label{tab:comparison-all-by-race}
\end{table}
 \end{small}
% latex table generated in R 3.5.1 by xtable 1.8-4 package
% Wed Dec  4 11:33:45 2019
\begin{table}[ht]
\centering
\begin{tabular}{lrrrr}
 group & exclusions & bump-ups & nvca & rec \\ 
  \hline
non-black & 18.1 & 17.0 & 8.2 & 27.5 \\ 
  black & 24.5 & 16.9 & 14.5 & 27.2 \\ 
   \hline
\end{tabular}
\caption{Percent of cases by defendant race with a charge-based exclusion, charge-based bump-up, NVCA flag, or higher recommendations due to unsubstantiated booking charges.} 
\label{tab:all-comparison-prop-race}
\end{table}

To understand how this seemingly paradoxical result is possible, we must first recognize that there are instances where an individual can have an ``unfair'' charge-based exclusion or bump-up that does not translate to an ``unfair'' recommendation. Recall that each of these charge-based components results in an increase to the recommended level of supervision above and beyond the initial recommendation. 

Consider, for example, an individual whose initial recommendation is the highest level and who has exclusion or bump-up charges at booking that they are not convicted of. This individual would be classified as having an unfair exclusion or bump-up. However, because their initial recommendation was maximal, whether we calculate the PSA using the booking charges (which would include an exclusion or a bump-up) or we calculate it using only the conviction charges (which would not include an exclusion or bump-up), the recommendation is the same. In the former case, the initial recommendation was maximal and applying the exclusion or bump-up did not increase the recommendation, as it could not further increase. In the latter case, we do not apply the exclusion or bump-up, and the recommendation is still the highest category.  Thus, even if exclusion or bump-up booking charges are unsubstantiated, the ultimate recommendation does not change based on those charges for people whose initial recommendation is the highest level. Individuals in this category contribute to the disparity shown under exclusions and bump-ups in Tables \ref{tab:comparison-all-by-race} and  \ref{tab:all-comparison-prop-race} and do not contribute to any difference in recommendations.

Similarly, consider a second scenario where an individual has an initial recommendation of SFPDP-ACM, the second highest level of supervision. If this person is booked under an exclusion charge that is reduced to a bump-up charge that they are convicted of, in both cases, the final recommendation will be the highest level of supervision. To break this down further,  under the booking charges, they receive an exclusion and are automatically moved to the highest category, Release Not Recommended. Under the conviction charges, they receive a charge-based bump-up, which because they began in the second-highest category, also results in a Release Not Recommended recommendation. Thus, under both the booking charges and the conviction charges, their recommendation will be the same, though they will still be classified as having had an unfair exclusion.

\begin{figure}[h]
\centering
\includegraphics[width = 3in]{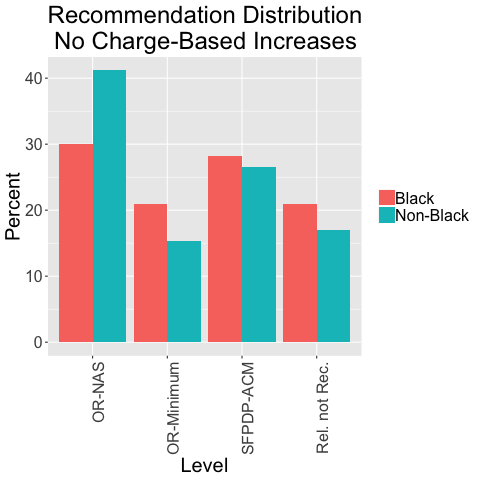}
\caption{\label{fig:pre-step-dist} Distribution of initial recommendation, disaggregated into Black and non-Black people.}
\end{figure}

Both scenarios where unfair exclusion charges do not materialize into unfair recommendations are only possible when the individual has an initial recommendation that is either the highest category or the second highest category. In the population examined here, the distribution of initial scores was shifted higher for the Black individuals than the non-Black individuals. See Figure \ref{fig:pre-step-dist}, which shows the distribution of initial scores broken down into Black and non-Black people. Depending on the definition of fairness adopted, this group-wise distributional difference may itself be indicative of unfairness in the model. However, because our goal is to study the effect of overbooking in isolation, we do not further delve into this other than to note this disparity in the rate at which Black versus non-back people are recommended for pre-trial detention. Regardless, because Black defendants were more likely to fall into the highest or second highest category before any charge-based amendments were made, Black defendants who had unfair charge-based exclusions were more likely to not have those unfair charge-based exclusions impact their final recommendation. 

It is important to note that this conclusion only holds for this particular DMF, the matrix that translates raw predictions of FTA and NCA into initial recommendations. In a jurisdiction with a different DMF under which fewer combinations of risk scores correspond to the highest or second-highest level of supervision, it is possible that the racial disparities in unwarranted charge-based exclusions, bump-ups, and NVCA flags might, in fact, translate to disparities in the recommendations as well. Thus, jurisdictions seeking to amend their DMF (or similarly, their release recommendation matrices) should be aware of this possibility: changes to the release recommendations that are intended to result in a greater rate of release may introduce a racial skew in the impact of overbooking where one previously did not exist.

\section{Limitations}\label{sec:Limitations}
Perhaps most obviously, one limitation of this analysis is that it was done using a fairly small sample size with a limited scope of the cases it covers. This sample may not include all of the PSAs administered during the time period.  In order to generalize the findings here, one would need to repeat the analysis using data from a broader geographic range and preferably across a longer time horizon.  

Additionally, as with many endeavors to measure criminality, our measure is imperfect. As discussed in the introduction,  higher booking charges may translate to higher conviction charges, as later stages of the criminal justice process anchoring to the booking charges. This would cause our analysis to understate the impact of overbooking. The other side of the argument is that because we have excluded conviction charges that are associated with other cases, even when the individual pleaded guilty to those charges as part of a joint deal covering both the case we consider and the case containing the conviction charges, we are understating the extent of guilty charges that ought to count. This would lead to us overstating the extent of overbooking. In the end, there is no perfect measure of criminal behavior, and we believe the measure we have chosen is  reasonable.

Finally, we have analyzed the risk assessment in isolation, rather than taking a more holistic approach to analyzing its role within the larger system. For example, we have presented no analysis of the impact of overbooking on judicial decision-making nor analyzed the effect on downstream outcomes like recidivism. In order to understand the real world consequences of overbooking on the individuals evaluated by the risk assessment, more investigation is needed. In this vein, the observed lack of racial disparity in the extent to which overbooking impacts the tool's final recommendation should not be taken as definitive proof that overbooking does not have a racially disparate impact on judicial decision-making. Each component of the PSA we have evaluated is displayed on the sheet available to judges. It is possible that the presence of a charge-based exclusion, bump-up, or NVCA flag influences the judge's decision-making independent of the tool's final recommendation. If this is so, then the finding that overbooking impacted charge-based exclusions at a higher rate for Black than non-Black individuals, for example, might actually result in a meaningful difference in terms of the impact of overbooking on the judge's decision. This despite the fact that we observed no difference between race groups in the impact of overbooking on the final recommendation of the tool. Without data on what decisions judges made when presented with these evaluations, we cannot say what impact overbooking had on the decisions at an aggregate level or disaggregated by race.

 \section{Conclusion}\label{sec:conclusion}
We have analyzed the effect of overbooking on a popular risk assessment tool, Arnold Ventures' PSA. We have found that for around 27\% of the cases analyzed, charges for which the person was not ultimately convicted caused the tool to issue a recommendation for pre-trial supervision that was more restrictive than would have been issued had such charges not been included as inputs to the model.  At a more granular level, we have found that a significant portion of the population that was evaluated by the tool, between 10- 20\%, received charge-based exclusions, charge-based bump-ups, and NVCA flags also based upon charges that were ultimately unsubstantiated by the courts. Disaggregating the analysis by race shows that while Black individuals received unwarranted charge-based exclusions and NVCA flags at a higher rate than non-Black individuals, they did not receive increased recommendations at a substantially higher rate due to the fact that Black individuals were more likely to be classified in the higher risk groups even before charge-based increases are applied. This  finding  in and of itself may be worrisome to those who hope that risk assessment will close the racial gap in criminal justice outcomes.  

This work also naturally raises the question of how one might protect against charges that will ultimately not result in convictions impacting an individual's pre-trial release recommendation. It is not possible to wait until a case is resolved to use only conviction charges to compute a recommendation for pre-trial release conditions, as by definition, by that time the pre-trial period has come to a close.  One possible mitigation is to have a defendant advocate present earlier in the process to help catch charges for which the early evidence is lacking. This may be difficult in practice, as the police report is typically not written at the time the PSA is administered. Without the police report, a defendant advocate may not have sufficient information to assess the strength of the evidence for each of the charges. %Jurisdictions that see large rates of overbooking and are keen on using a risk assessment might also consider using alternative risk assessment tools that do not include un-vetted charge information. 

This work also reveals one possibility for gaming the risk assessment. In theory, in cases where there is ambiguity as to which charges are appropriate for a given observed behavior, a booking officer could opt to book under more serious charges with the intention of inducing a more restrictive detention decision. This could be monitored by tracking the ratio of similar higher and lower charges over time and across protected group status. A heightened ratio may indicate strategic behavior intended to prompt higher recommendations by the tool. 

This possibility points to the fragility of such models under highly discretionary inputs. In our view, this raises questions about the appropriateness of their use in high stakes settings without adequate controls and accountability for the inputs. In particular, we believe this reveals the importance of increased accountability and feedback for officers who systematically overbook. This is especially important if booking charges are to be legitimized by their use as inputs to the tool, the output of which--- though directly dependent on booking charges--- may not be viewed with the same level of skepticism as the charges themselves might be. One potentially revealing line of future research would be to disaggregate the impact of overbooking by booking officer as a means to determine if particular officers' booking decisions are routinely unfairly impacting the tool's pre-trial supervision recommendations. Holding such individuals accountable for their booking decisions may encourage more conservative decisions around booking arrested people on heightened charges, reducing the impact of overbooking even beyond the risk assessment. 

Finally,  while judges have the power to set more restrictive conditions of release for people booked on certain serious charges,  these types of charges do not necessarily correlate with an increased risk of undesirable outcomes. Embedding charge-based considerations within the final recommendation may communicate to judges that merely having been booked on such charges increases an individual's empirical risk level when, in fact, such overrides are an instantiation of a policy driven by politics rather than being directly tied to an individual's risk. Given the large impact of charging decisions on the output of the tool, it is critical that judges understand this distinction in order to properly discount increased recommendations that are based on questionably or weakly supported charges. 
 
\appendix
\section{De-duplication and record-linkage} \label{sec:matching}
To operationalize the definition of ``fair" booking charges we have adopted, we must track down the outcome of the case associated with each PSA. This requires linking the court records, which contain outcome status by charge, to the PSA data. In this process, we only consider records for which the full PSA was completed. We exclude any PSA records missing inputs for any of the PSA components, such as arrest date and any of the PSA predictions (FTA, NCA, NCVA). We also check that each complete record in the PSA data is unique in a process called de-duplication. We define two records as pertaining to the same incident if both records are for the same individual (have the same SFID), were recorded on the same day, and list the same charge(s). If there are multiple records of the same individual on the same day with the same charge(s), we retain only one copy. %The logic here is that it is unlikely that an individual would be arrested, released, and re-arrested on the same day for the same charge(s). Duplicates in the dataset could reasonably have arisen by multiple people scanning and saving the same PSA file and saving it in the archive under different file names. They also could have arisen upon data entry if the person manually entering the data from the scanned documents accidentally entered the same file multiple times. 
%We  drop mnum$num_dropped_dups records as duplicates. It is worth noting that these mnum$num_dropped_dups correspond to only mnum$num_unique_date_and_charge unique arrests, meaning there many cases where there were several duplicate PSAs in the data for the same arrest event. This leaves mnum$psa_recs_after_drop_dupes PSA records to match. 

To match the PSA data to the court data, for each PSA record, we first identify all court numbers (a unique identifier of a case--- one individual can have multiple court numbers) that list the same SFID and an arrest date within one day before to two days after that listed in the PSA record. This determines the initial set of potential matches. For the vast majority of PSAs, there is only one case in this set, and it is declared the match. If there are multiple court numbers that meet these criteria, we select the court number(s) with the charges that most closely match those listed on the PSA using the following procedure.

We determine whether each potential match in the court records contains the charge code listed as the top charge on the PSA form. If only one of the potential matches does, this is declared the match. If multiple potential matches contain the charge code listed first on the PSA, we repeat this using the second listed charge code if one exists. In most cases, there is only one court number that matches on both first and second charges and it is declared the match. If there are multiple, we declare all of them to be matches, and all charges associated with the matched court numbers are included in our calculations.  %This results in 1916 matches of the 1967 PSA records we set out to find matches for. 

For those records for which this process found no match, we manually search through the court records to identify an appropriate match. There are several idiosyncratic reasons why matches were not found using our automated procedure. For example, in one case, the day and month of arrest had been transposed in the PSA data relative to the court data, i.e. they used different date formats. We consider this and other similar errors to be simple recording errors, and we manually determined a match on a case-by-case basis. In the end, despite our best efforts to find a match in the court records for every PSA, there were 31 PSA records for which we were unable to definitively find a match, and these were dropped from the analysis.

\section{Validation of our reproduction of the PSA}\label{sec:validation}
Using the matched data, we  confirm that our code is able to reproduce the results of the PSA based on the booking charges listed in the court data. %If it does, we can be confident that when we apply the code to counterfactual scenarios in which only ``fair" booking charges are used, the output of the code will be the same as what the output would have been had the ``fair'' booking charges been used to calculate the PSA. 
We wrote a function that takes as input the PSA's NCA and FTA predictions (which are not a function of booking charge so will not be sensitive to changes to this input), the factors used to calculate the NVCA flag (which is dependent on booking charges), and the list of booking charges as taken from the court records. Sometimes, this differs from the booking charges listed on the PSA form. As output, it returns our calculations for the NVCA flag, an indicator of a charge-based exclusion, an indicator of a charge-based bump-up, and the final recommendation.%  We again note that because the goal of this analysis is to isolate the effect of the booking charges, we do not explore changing the NCA and FTA predictions, as these are not dependent on the booking charge. %We take these as fixed and assess how the input charges affect the output. 

\subsection{NVCA flag}

We follow the procedure to calculate the NVCA flag as outlined in the PSA Portal. One of the inputs to the NVCA flag is whether the booking charge is violent. 
%According to Arnold Ventures's PSA information portal, the NVCA flag is calculated by assigning a point value to each of several factors: current violent offense (2 points), current violent offense and 20 years old or younger (1 point), pending charge at the time of the offense (1 point), any prior convictions (1 point), and prior violent conviction (1 point for one violent conviction; 2 points for two or three violent convictions; 3 points for more than three violent convictions). Note that the first two factors depend on the booking charges used to calculate them. If the point values sum to greater than four,  he NVCA flag is one. Otherwise, it is zero.
For each PSA in our dataset, we compare the booking charges listed in the court records to the California PSA List of Violent Charges. If any of the booking charges are present on the list, we calculate the NVCA flag as though the current offense is violent. All other  non-charge-based inputs to the NVCA flag are taken directly from the original PSA form. 
%the NVCA flag is the individual is assigned two points. If the any of the charges are on the list and the individual's age is listed as 20 or less on the date of arrest, they receive an additional one point. We take the other point values for all other factors directly from the original PSA form. Individuals whose re-calculated NVCA risk score is four or greater are assigned a reproduced NVCA flag of one; those whose NVCA risk score is less than four receive a reproduced NVCA flag of zero. We then compare our reproduction of the NVCA flag to that listed on the PSA form.\footnote{The PSA forms do not include a raw NVCA score. They only indicate whether the NVCA flag has been triggered. Thus, we cannot compare our calculation of the raw score to that assigned by the administrator of the PSA.}
We find that our calculation of the NVCA flag agrees with that listed on the PSA form for 99.3\% of the cases in our dataset. 
 
\subsection{Charge-based exclusions and bump-ups}
%We calculate charge-based exclusions and bump-ups in accordance with the PSA as outlined in Section \ref{sec:PSA} using the booking charges as listed in the court records as inputs.  With regard to charge-based exclusions, we find that our reproduction is in agreement with the original PSA data for mnum$step2_agree\% of the records. 
%
%In the PSA, calculating the charge-based bump-up is not necessary if an exclusion is determined because the recommended level of supervision cannot further be increased. For this reason, we compare our reproduction of charge-based bump-ups to the PSA data only for those records for which a charge-based exclusion was not indicated in the original PSA data.\footnote{Based on a manual review, it seems that it was inconsistent whether charge-based bump-ups were calculated if an exclusion was already indicated.} We find that our reproduction is in agreement with the original administration of the PSA mnum$step4_agree\% of these records.  Due to the high level of agreement between our reproduction of the PSA's charge-based exclusions and bump-ups and those recorded in the original PSA data, we are satisfied that our code is adequately replicating this portion of the PSA. 
%
In cases where there was a discrepancy between our code and what appears on the human-administered PSA, we did a manual review. In most cases, the discrepancy was due to a slightly different set of booking charges listed in the court records than were listed on the PSA form. In other cases, we believe it was due to inconsistency in how the PSA administrators handled ambiguous instructions. For example, recall that a charge-based bump-up is indicated if the booking charge ``involved the {\underline{Use} of a weapon, caustic chemical, flammable substance, or explosive." There were some cases in the PSA data where an individual received a bump-up based on a charge of  wielding an imitation firearm (417.4 PC). Other cases that included this charge did not receive a bump-up, which we believe indicates some inconsistency in how bump-ups were applied with respect to this charge due to the ambiguity around whether an imitation firearm should be considered a firearm. Similar inconsistencies were noted around the interpretation of whether carrying a loaded firearm (25850(A) PC) counted as  ``\underline{Use} of a weapon" for the purposes of a charge-based bump-ups.

\subsection{Final recommendation}
Finally, we compare our calculation of the final recommendation to the recommendation given in the PSA data. We find an agreement of 97.4\%, which is slightly lower than for the other components. Further inspection reveals that the majority of the disagreements occur in the case when the FTA prediction was 5 and the NCA prediction was 4. This combination of NCA and FTA scores corresponds to the cell in the decision-making framework (see Figure \ref{fig:dmf}) where additional decision rules are applied, presumably by a human. For cases that have this combination of NCA and FTA scores, if any of the booked offenses are a felony or a violent misdemeanor as indicated by the PSA Violent Offense List, the recommendation is ``Release Not Recommended." Otherwise, the recommendation is ``SFPDP-ACM."  

We also manually reviewed several of these disagreements. In many cases of disagreement, there was a felony charge that did not appear on the violent offense list. Under the rule as we interpret it, any felony leads to the increase in level. We suspect that this rule was sometimes interpreted to mean that only violent felonies triggered the additional increase. In any case, for those cases that do not fall into this cell of the DMF, regarding the final recommendation, our reproduction is in agreement  with the PSA data 99.5\% of the time.

\section{Violent charge list}
The California PSA violent charge list is the list of charges used by the PSA to define violence. This list contains over 200 charge codes, so we do not reproduce the whole list here. Charge codes are separated on the list into several categories. For each category of charge that appears on the list, we give some example charges and their associated charge codes. We have tried to select charges that are representative of those that appear in each category without being redundant.

\begin{small}
\begin{table*}
\begin{tabular}{L{5.5cm}|L{5.5cm}|L{3cm} }
%\begin{tabular}{l|l|l}
\hline
Category & Example Charge & Example Code    \\  \hline \hline
 Other offenses against public justice    &    Resist Police Officer: Cause Death/Serious Bodily Injury; Threatening witnesses, victims, or informants; Advocacy to kill or injure peace officer            & 148.10(A) PC F;   140(a) M; 151      \\ \hline
Homicide & Murder first degree; Gross vehicular manslaughter while intoxicated; Solicit to commit murder & 187(A) PC F 1; 191.5(A) PC F; 653F(B) PC F\\\hline
Mayhem & Mayhem; Torture & 203PC F; 206PC F \\\hline
Kidnapping & Kidnapping; False imprisonment of a hostage  & 207(A) PC F; 210.5 PC F \\\hline
Robbery & Robbery: first degree; Carjacking & 211 PC F 1; 215(A) PC F \\\hline
Attempts to kill & Assault on a public official; Attempted murder of public official; Obstructing railroad track, punishment; Throwing missile at common carrier with bodily harm & 217.1(A) PC F; 217.1(B) PC F; 218.1 F; 219.1 F \\\hline
  Assaults with intent to commit felony, not murder & Assault with intent to commit a felony; Assault to commit a felony during the commission of first degree burglary & 220 (A)(1) PC F; 220(B) PC  F \\\hline
  False imprisonment and human trafficking & False imprisonment; Human trafficking & 236 PC M; 236.1 \\\hline
Assault and battery & Assault; Assault on a peace officer of a school district; Assault on a highway worker; Battery on school employee;  Battery against police, emergency personnel, etc.; Assault w/firearm on person; Shooting at an  inhabited dwelling, vehicle, etc. & 240 PC M; 241.4 M  F; 241.5 M; 241.6 PC M; 243(B) PC M; 245(A)(2) PC; 246PC F\\\hline
  Rape, abduction, carnal abuse of children & Rape: victim incapable of consent; Sex intercourse w/a minor less than18; Rape spouse by force/etc; Attempted lewd acts/w/child und 14yrs & 261(A)(1) PC F;  261.5(A) PC F; 262(A)(1) PC F; 664/288 (A) PC F \\ \hline
 Abandonment and neglect of children & Willful cruelty to child; Injuring a spouse, cohabitant, fianc\'e, boyfriend, girlfriend or child’s parent; Inflict injury upon child & 273 A(B) PC M; 273.5 (A) PC M; 273 D(A) PC M \\\hline
 Bigamy, incest, and crimes against nature & Incest; Sodomy: person under 18;  Lewd/lascivious acts on dependent adult w/force; Oral copulation; Harmful material sent w/intent to seduce minor; Sexual penetration w/foreign object: victim drugged; Continuous sexual abuse of child & 285 PC F; 286(B)(1) PC F; 288 (B)(2) PC F; 288 A(A) PC F; 288.2 (A) PC F; 289 (E) PC F; 288.5 (A) PC F \\\hline
 Other injuries to person & Poisoning & 347(A) PC F\\\hline
 Crimes against elders, dependent adults, persons with disabilities & Cause harm/death elder dependent adult; Elder abuse: victim 70/or older & 368(B)(1) PC M; 368(b)(2)(B) PC F \\\hline
 Crimes against the public peace & Rioting; Exhibit firearm. Drawing, exhibiting, or using a firearm & 404(A) PC M; 417 (b) PC M \\\hline
Crimes and penalties & Violate civil rights by force/threat & 422.6(A) PC M \\\hline
Arson & Arson causing great bodily injury; Aggravated arson; Causing fire of inhabited structure/property & 451(A) PC F; 451.5 (A) PC F; 452 (B) PC F\\\hline
Offenses by prisoners & Assault by a life prisoner; Escape from custody by force and violence & 4500 F; 4530(A) PC F\\\hline
 Prevention and abatement of unlawful activity & Use of a weapon of mass destruction causing death & 11418 (B)(2) PC F\\\hline
  Destructive devices and explosives generally & Use of destructive device and explosive to injure/destroy; Explosion causing death & 18740 F; 18755 (a) F\\\hline \hline
\end{tabular}
\caption{\label{tab:violent-offenses} California PSA Violent Offenses List: representative charges by category}
\end{table*}
\end{small}

\section{Consistency of Racial Classification in Court Data}
In order to break down this analysis by the race of the arrested person, we first explore how consistently individuals were assigned to each race category. The court records contain a field called race which takes values B, C,  F, H, I, J, U, W, and in some cases it is missing. We believe these stand for Black, Chinese, Filipino,  Hispanic, Indian, Japanese, Other, Unknown, and White, respectively. Some of these categories (F, I, J) were very rarely used.

To calculate our measure of consistency, we first restrict the analysis to all individuals who had more than one arrest record in the dataset. For each race category (shown in rows of Table \ref{tab:race-consistency}), we filter the data to include only individuals who are listed as that race at least once in the data. Using this subset, for each individual we calculate the percent of their records that were classified as each possible race. We then take an average across all individuals.  This serves as our measure of how consistently the individuals are categorized into each group. The $ij$th entry of Table \ref{tab:race-consistency} is then the average percent of records that were classified as the $j$th race category among all individuals who were classified as the $i$th race category at least once. The rows do not sum to 100 as one might expect, due to the omission of rows and columns corresponding to rare categories and missing values. Looking at the row labeled as $B$, we see that for an individual who was categorized as $B$ at least once, on average 98.4\% of their records were labeled as $B$ and 0.4\% off their records were labeled as $W$. This shows a high level of consistency in designating people as $B$. However, for a person labeled as $H$ at least once, 35.7\% of their records were indicated a $W$, leading us to conclude there is a reasonably high degree of variability in how people who may be considered $H$ are classified. 
By looking at the diagonal of this table,  we see that the only category that is highly consistent under this measure is $B$ with approximately 98\% agreement, and so for the analysis presented here, the only race categories we consider are $B$ and non- $B$, which we assume to be Black and non-Black. 
% latex table generated in R 3.5.1 by xtable 1.8-4 package
% Mon Jun  3 17:24:12 2019
\begin{table}[ht]
\centering
\begin{tabular}{lrrrrrr}
  \hline
Race Designation & B & C & H & O & U & W \\ 
  \hline
B & 98.40 & 0.10 & 0.10 & 0.20 & 0.30 & 0.40 \\ 
  C & 0.50 & 85.60 & 0.00 & 5.90 & 1.60 & 3.20 \\ 
  H & 0.50 & 0.00 & 48.40 & 4.20 & 2.80 & 35.70 \\ 
  O & 2.90 & 14.90 & 2.90 & 46.00 & 3.40 & 25.90 \\ 
  U & 7.70 & 2.60 & 9.00 & 3.80 & 53.80 & 16.70 \\ 
  W & 0.40 & 0.10 & 2.90 & 1.90 & 0.40 & 91.60 \\ 
   \hline
\end{tabular}
\caption{The expected rate per 100 records at which individuals which each race designation and more than one record were classified as each possible designation (columns). For example, for individuals who were classified as U at least once and had more than one record available, on average those individuals had  8  \% of their records listed as B.} 
\label{tab:race-consistency}
\end{table}

\begin{acks}
We are grateful to Chelsea Barabas, Logan Koepke,  Alicia Solow-Niederman, and David Robinson for helpful comments on an early draft. This work was supported by The Ethics and Governance of AI Initiative The MacArthur Foundation, and the Ford Foundation.

\end{acks}

\bibliographystyle{ACM-Reference-Format}
\bibliography{HRDAG_Bib_files.bib}


\begin{thebibliography}{21}

%%% the Web address.

\ifx \showCODEN    \undefined \def \showCODEN     #1{\unskip}     \fi
\ifx \showDOI      \undefined \def \showDOI       #1{#1}\fi
\ifx \showISBNx    \undefined \def \showISBNx     #1{\unskip}     \fi
\ifx \showISBNxiii \undefined \def \showISBNxiii  #1{\unskip}     \fi
\ifx \showISSN     \undefined \def \showISSN      #1{\unskip}     \fi
\ifx \showLCCN     \undefined \def \showLCCN      #1{\unskip}     \fi
\ifx \shownote     \undefined \def \shownote      #1{#1}          \fi
\ifx \showarticletitle \undefined \def \showarticletitle #1{#1}   \fi
\ifx \showURL      \undefined \def \showURL       {\relax}        \fi
% invisible to TeX
\providecommand\bibfield[2]{#2}
\providecommand\bibinfo[2]{#2}
\providecommand\natexlab[1]{#1}
\providecommand\showeprint[2][]{arXiv:#2}

\bibitem[\protect\citeauthoryear{Angwin, Larson, Mattu, and Kirchner}{Angwin
  et~al\mbox{.}}{2016}]%
        {angwin2016machine}
\bibfield{author}{\bibinfo{person}{Julia Angwin}, \bibinfo{person}{Jeff
  Larson}, \bibinfo{person}{Surya Mattu}, {and} \bibinfo{person}{Lauren
  Kirchner}.} \bibinfo{year}{2016}\natexlab{}.
\newblock \showarticletitle{Machine bias}.
\newblock \bibinfo{journal}{\emph{ProPublica, May}}  \bibinfo{volume}{23}
  (\bibinfo{year}{2016}).
\newblock


\bibitem[\protect\citeauthoryear{Berk, Heidari, Jabbari, Kearns, and Roth}{Berk
  et~al\mbox{.}}{2018}]%
        {berk2018fairness}
\bibfield{author}{\bibinfo{person}{Richard Berk}, \bibinfo{person}{Hoda
  Heidari}, \bibinfo{person}{Shahin Jabbari}, \bibinfo{person}{Michael Kearns},
  {and} \bibinfo{person}{Aaron Roth}.} \bibinfo{year}{2018}\natexlab{}.
\newblock \showarticletitle{Fairness in criminal justice risk assessments: The
  state of the art}.
\newblock \bibinfo{journal}{\emph{Sociological Methods \& Research}}
  (\bibinfo{year}{2018}).
\newblock


\bibitem[\protect\citeauthoryear{Chouldechova}{Chouldechova}{2017}]%
        {chouldechova2017fair}
\bibfield{author}{\bibinfo{person}{Alexandra Chouldechova}.}
  \bibinfo{year}{2017}\natexlab{}.
\newblock \showarticletitle{Fair prediction with disparate impact: A study of
  bias in recidivism prediction instruments}.
\newblock \bibinfo{journal}{\emph{Big Data}} \bibinfo{volume}{5},
  \bibinfo{number}{2} (\bibinfo{year}{2017}), \bibinfo{pages}{153--163}.
\newblock


\bibitem[\protect\citeauthoryear{DeMichele, Baumgartner, Wenger, Barrick,
  Comfort, and Misra}{DeMichele et~al\mbox{.}}{2018}]%
        {demichele2018public}
\bibfield{author}{\bibinfo{person}{Matthew DeMichele}, \bibinfo{person}{Peter
  Baumgartner}, \bibinfo{person}{Michael Wenger}, \bibinfo{person}{Kelle
  Barrick}, \bibinfo{person}{Megan Comfort}, {and} \bibinfo{person}{Shilpi
  Misra}.} \bibinfo{year}{2018}\natexlab{}.
\newblock \showarticletitle{The Public Safety Assessment: A Re-validation and
  Assessment of Predictive Utility and Differential Prediction by Race and
  Gender in Kentucky}.
\newblock  (\bibinfo{year}{2018}).
\newblock


\bibitem[\protect\citeauthoryear{Dieterich, Mendoza, and Brennan}{Dieterich
  et~al\mbox{.}}{2016}]%
        {dieterich2016compas}
\bibfield{author}{\bibinfo{person}{William Dieterich},
  \bibinfo{person}{Christina Mendoza}, {and} \bibinfo{person}{Tim Brennan}.}
  \bibinfo{year}{2016}\natexlab{}.
\newblock \showarticletitle{COMPAS risk scales: Demonstrating accuracy equity
  and predictive parity}.
\newblock \bibinfo{journal}{\emph{Northpoint Inc}} (\bibinfo{year}{2016}).
\newblock


\bibitem[\protect\citeauthoryear{Dobbie, Goldin, and Yang}{Dobbie
  et~al\mbox{.}}{2018}]%
        {dobbie2018effects}
\bibfield{author}{\bibinfo{person}{Will Dobbie}, \bibinfo{person}{Jacob
  Goldin}, {and} \bibinfo{person}{Crystal~S Yang}.}
  \bibinfo{year}{2018}\natexlab{}.
\newblock \showarticletitle{The effects of pretrial detention on conviction,
  future crime, and employment: Evidence from randomly assigned judges}.
\newblock \bibinfo{journal}{\emph{American Economic Review}}
  \bibinfo{volume}{108}, \bibinfo{number}{2} (\bibinfo{year}{2018}),
  \bibinfo{pages}{201--40}.
\newblock


\bibitem[\protect\citeauthoryear{Eckhouse, Lum, Conti-Cook, and
  Ciccolini}{Eckhouse et~al\mbox{.}}{2019}]%
        {eckhouse2019layers}
\bibfield{author}{\bibinfo{person}{Laurel Eckhouse}, \bibinfo{person}{Kristian
  Lum}, \bibinfo{person}{Cynthia Conti-Cook}, {and} \bibinfo{person}{Julie
  Ciccolini}.} \bibinfo{year}{2019}\natexlab{}.
\newblock \showarticletitle{Layers of bias: A unified approach for
  understanding problems with risk assessment}.
\newblock \bibinfo{journal}{\emph{Criminal Justice and Behavior}}
  \bibinfo{volume}{46}, \bibinfo{number}{2} (\bibinfo{year}{2019}),
  \bibinfo{pages}{185--209}.
\newblock


\bibitem[\protect\citeauthoryear{Gupta, Hansman, and Frenchman}{Gupta
  et~al\mbox{.}}{2016}]%
        {gupta2016heavy}
\bibfield{author}{\bibinfo{person}{Arpit Gupta}, \bibinfo{person}{Christopher
  Hansman}, {and} \bibinfo{person}{Ethan Frenchman}.}
  \bibinfo{year}{2016}\natexlab{}.
\newblock \showarticletitle{The heavy costs of high bail: Evidence from judge
  randomization}.
\newblock \bibinfo{journal}{\emph{The Journal of Legal Studies}}
  \bibinfo{volume}{45}, \bibinfo{number}{2} (\bibinfo{year}{2016}),
  \bibinfo{pages}{471--505}.
\newblock


\bibitem[\protect\citeauthoryear{Harcourt}{Harcourt}{2015}]%
        {harcourt2015risk}
\bibfield{author}{\bibinfo{person}{Bernard~E Harcourt}.}
  \bibinfo{year}{2015}\natexlab{}.
\newblock \showarticletitle{Risk as a proxy for race: The dangers of risk
  assessment}.
\newblock \bibinfo{journal}{\emph{Federal Sentencing Reporter}}
  \bibinfo{volume}{27}, \bibinfo{number}{4} (\bibinfo{year}{2015}),
  \bibinfo{pages}{237--243}.
\newblock


\bibitem[\protect\citeauthoryear{Johndrow, Lum, et~al\mbox{.}}{Johndrow
  et~al\mbox{.}}{2019}]%
        {johndrow2019algorithm}
\bibfield{author}{\bibinfo{person}{James~E Johndrow}, \bibinfo{person}{Kristian
  Lum}, {et~al\mbox{.}}} \bibinfo{year}{2019}\natexlab{}.
\newblock \showarticletitle{An algorithm for removing sensitive information:
  application to race-independent recidivism prediction}.
\newblock \bibinfo{journal}{\emph{The Annals of Applied Statistics}}
  \bibinfo{volume}{13}, \bibinfo{number}{1} (\bibinfo{year}{2019}),
  \bibinfo{pages}{189--220}.
\newblock


\bibitem[\protect\citeauthoryear{Kleinberg}{Kleinberg}{2018}]%
        {Kleinberg:2018:ITA:3219617.3219634}
\bibfield{author}{\bibinfo{person}{Jon Kleinberg}.}
  \bibinfo{year}{2018}\natexlab{}.
\newblock \showarticletitle{Inherent Trade-Offs in Algorithmic Fairness}. In
  \bibinfo{booktitle}{\emph{Abstracts of the 2018 ACM International Conference
  on Measurement and Modeling of Computer Systems}}
  \emph{(\bibinfo{series}{SIGMETRICS '18})}. \bibinfo{publisher}{ACM},
  \bibinfo{address}{New York, NY, USA}, \bibinfo{pages}{40--40}.
\newblock


\bibitem[\protect\citeauthoryear{Kurlychek and Johnson}{Kurlychek and
  Johnson}{2019}]%
        {kurlychek2019cumulative}
\bibfield{author}{\bibinfo{person}{Megan~C Kurlychek} {and}
  \bibinfo{person}{Brian~D Johnson}.} \bibinfo{year}{2019}\natexlab{}.
\newblock \showarticletitle{Cumulative Disadvantage in the American Criminal
  Justice System}.
\newblock \bibinfo{journal}{\emph{Annual Review of Criminology}}
  \bibinfo{volume}{2} (\bibinfo{year}{2019}), \bibinfo{pages}{291--319}.
\newblock


\bibitem[\protect\citeauthoryear{Leslie and Pope}{Leslie and Pope}{2017}]%
        {leslie2017unintended}
\bibfield{author}{\bibinfo{person}{Emily Leslie} {and} \bibinfo{person}{Nolan~G
  Pope}.} \bibinfo{year}{2017}\natexlab{}.
\newblock \showarticletitle{The unintended impact of pretrial detention on case
  outcomes: Evidence from New York City arraignments}.
\newblock \bibinfo{journal}{\emph{The Journal of Law and Economics}}
  \bibinfo{volume}{60}, \bibinfo{number}{3} (\bibinfo{year}{2017}),
  \bibinfo{pages}{529--557}.
\newblock


\bibitem[\protect\citeauthoryear{Lum, Ma, and Baiocchi}{Lum
  et~al\mbox{.}}{2017}]%
        {lum2017causal}
\bibfield{author}{\bibinfo{person}{Kristian Lum}, \bibinfo{person}{Erwin Ma},
  {and} \bibinfo{person}{Mike Baiocchi}.} \bibinfo{year}{2017}\natexlab{}.
\newblock \showarticletitle{The causal impact of bail on case outcomes for
  indigent defendants in New York City}.
\newblock \bibinfo{journal}{\emph{Observational Studies}}  \bibinfo{volume}{3}
  (\bibinfo{year}{2017}), \bibinfo{pages}{39--64}.
\newblock


\bibitem[\protect\citeauthoryear{Mayson}{Mayson}{2019}]%
        {Mayson:2019aa}
\bibfield{author}{\bibinfo{person}{Sandra~G. Mayson}.}
  \bibinfo{year}{2019}\natexlab{}.
\newblock \showarticletitle{Bias In, Bias Out}.
\newblock \bibinfo{journal}{\emph{The Yale Law Journal}} \bibinfo{volume}{128},
  \bibinfo{number}{8} (\bibinfo{year}{2019}), \bibinfo{pages}{2122--2473}.
\newblock


\bibitem[\protect\citeauthoryear{Owens, Kerrison, and Santos Da~Silveira}{Owens
  et~al\mbox{.}}{2017}]%
        {owens2017examining}
\bibfield{author}{\bibinfo{person}{Emily Owens}, \bibinfo{person}{Erin~M
  Kerrison}, {and} \bibinfo{person}{B Santos Da~Silveira}.}
  \bibinfo{year}{2017}\natexlab{}.
\newblock \showarticletitle{Examining racial disparities in criminal case
  outcomes among indigent defendants in San Francisco}.
\newblock \bibinfo{journal}{\emph{Quattrone Center for the Fair Administration
  of Justice, University of Pennsylvania Law School}} (\bibinfo{year}{2017}).
\newblock


\bibitem[\protect\citeauthoryear{Redcross, Henderson, Miratrix, and
  Valentine}{Redcross et~al\mbox{.}}{2019}]%
        {Redcross:2019aa}
\bibfield{author}{\bibinfo{person}{Cindy Redcross}, \bibinfo{person}{Brit
  Henderson}, \bibinfo{person}{Luke Miratrix}, {and} \bibinfo{person}{Erin
  Valentine}.} \bibinfo{year}{2019}\natexlab{}.
\newblock \bibinfo{booktitle}{\emph{Evaluation of Pretrial Justice System
  Reforms that Use the Public Safety Assessment}}.
\newblock \bibinfo{type}{{T}echnical {R}eport}. \bibinfo{institution}{MDRC
  Center for Criminal Justice Research}.
\newblock


\bibitem[\protect\citeauthoryear{Skeem and Lowenkamp}{Skeem and
  Lowenkamp}{2016}]%
        {skeem2016risk}
\bibfield{author}{\bibinfo{person}{Jennifer~L Skeem} {and}
  \bibinfo{person}{Christopher~T Lowenkamp}.} \bibinfo{year}{2016}\natexlab{}.
\newblock \showarticletitle{Risk, race, and recidivism: Predictive bias and
  disparate impact}.
\newblock \bibinfo{journal}{\emph{Criminology}} \bibinfo{volume}{54},
  \bibinfo{number}{4} (\bibinfo{year}{2016}), \bibinfo{pages}{680--712}.
\newblock


\bibitem[\protect\citeauthoryear{Solow-Niederman, Choi, and den
  Broeck}{Solow-Niederman et~al\mbox{.}}{[n. d.]}]%
        {Solow-Niederman:2019aa}
\bibfield{author}{\bibinfo{person}{Alicia Solow-Niederman},
  \bibinfo{person}{YooJung Choi}, {and} \bibinfo{person}{Guy~Van den Broeck}.}
  \bibinfo{year}{[n. d.]}\natexlab{}.
\newblock \showarticletitle{The Institutional Life of Algorithmic Risk
  Assessment}.
\newblock \bibinfo{journal}{\emph{Berkeley Technology Law Journal}}
  (\bibinfo{year}{[n. d.]}).
\newblock


\bibitem[\protect\citeauthoryear{Stevenson}{Stevenson}{2018}]%
        {stevenson2018distortion}
\bibfield{author}{\bibinfo{person}{Megan~T Stevenson}.}
  \bibinfo{year}{2018}\natexlab{}.
\newblock \showarticletitle{Distortion of justice: How the inability to pay
  bail affects case outcomes}.
\newblock \bibinfo{journal}{\emph{The Journal of Law, Economics, and
  Organization}} \bibinfo{volume}{34}, \bibinfo{number}{4}
  (\bibinfo{year}{2018}), \bibinfo{pages}{511--542}.
\newblock


\bibitem[\protect\citeauthoryear{Ventures}{Ventures}{[n. d.]}]%
        {Ventures:aa}
\bibfield{author}{\bibinfo{person}{Arnold Ventures}.} \bibinfo{year}{[n.
  d.]}\natexlab{}.
\newblock \bibinfo{booktitle}{\emph{Guide to the Release Conditions Matrix}}.
\newblock


\end{thebibliography}

\end{document}